\documentclass[12pt,a4paper]{article}

\usepackage{a4wide}
\usepackage{amsmath}
\usepackage{amssymb}
\usepackage{color}
\numberwithin{equation}{section}

\usepackage{url}
\usepackage{hyperref}
\usepackage{graphicx}

\newcommand{\GeV}{\,\mathrm{GeV}}
\newcommand{\TeV}{\,\mathrm{TeV}}
\newcommand{\fb}{\mathrm{fb}}

\title{\bf Implications of QCD radiative corrections on
  high-$\mathbf{p_T}$ Higgs searches}
\author{Andrea Banfi$^1$, Juli\'an Cancino$^2$
  \\[0.5em]\normalsize
  $^1$ Albert-Ludwigs-Universit\"at Freiburg, Physikalisches Institut,
  D-79104 Freiburg, Germany.  \\[0.1em]\normalsize $^2$ Institute for
  Theoretical Physics, ETH Zurich, 8093 Zurich, Switzerland.}
\date{}

\begin{document}

\maketitle

\begin{abstract}
  We discuss the effect of next-to-leading order (NLO) QCD corrections
  to the Higgsstrahlung process, where the Higgs boson decays to
  bottom quarks, using a partonic-level fully differential code.
  First we evaluate the impact of initial- and final-state gluon
  radiation on the reconstruction of a mass peak with the fat-jet
  analysis in the boosted regime at the LHC with $\sqrt s = 14\TeV$ as
  proposed in Butterworth et al. (2008)~\cite{Butterworth:2008iy}.
  We then consider the current CMS search strategy for this channel and
  compare it to the fat-jet procedure at the LHC with $\sqrt s =
  8\TeV$. Both studies show that final-state QCD radiation has a sizeable
  effect and should be taken properly into account.
\end{abstract}

\section{Introduction}
\label{sec:intro}

Although the LHC has started its operations only a couple of years ago
and at half the design energy, it has already provided plenty of
information on the existence of a Standard Model (SM) Higgs
boson. With the $5\,\mathrm{fb}^{-1}$ luminosity collected with $\sqrt
s = 7\TeV$ at the end of 2011, ATLAS was able to exclude the presence
of a SM Higgs boson in the range $133\GeV < m_H < 230\GeV$ and
$260\GeV < m_H < 437\GeV$~\cite{:2012si}, and CMS in the range
$129\GeV < m_H < 525\GeV$~\cite{Chatrchyan:2012tx}, in both cases at
99\% confidence level. Furthermore, adding $6\,\mathrm{fb}^{-1}$ of
the 2012 run at $\sqrt s = 8\TeV$, it has been recently possible to
discover a new boson with mass around $125\GeV$~\cite{:2012gk,:2012gu}. It
remains to establish whether this boson is indeed the SM
Higgs by studying in detail all its decay modes. A light Higgs boson
directly produced in gluon-gluon fusion (the process giving the
largest cross section) decays predominantly into a $b \bar b$ pair,
where the signal is overwhelmed by the huge QCD dijet background. This
is why the decay modes that have led to the discovery are those who do
not involve hadronic final states, like $H \to \gamma
\gamma$~\cite{:2012sk,Chatrchyan:2012tw}, $H \to W
W$~\cite{Aad:2012sc,Chatrchyan:2012ty} or $H\to
ZZ$~\cite{:2012sm,Chatrchyan:2012dg,Chatrchyan:2012ft}. Although they
are suppressed with respect to the dominant $b\bar b $ mode, it is
still possible to extract a signal from the background. There is yet
another possibility to exploit the $b \bar b$ decay of the Higgs
boson, namely when it is produced in association with a vector boson
$V$ ($W$ or $Z$): the Higgsstrahlung process. In this case there are various
possibilities to disentangle the signal over the large $V b\bar b $
background, some of which have been already used at the
LHC~\cite{Chatrchyan:2012ww,:2012zf}. Among them, one of the most
promising strategies makes use of the fact that at the LHC, especially at
$\sqrt s = 14\TeV$, it is possible to produce particles with
transverse momenta well above their masses. It is the so-called
``boosted'' regime, in which one can reconstruct heavy particles
decaying hadronically, because their decay products are likely to fall
inside one jet with a large radius, a.k.a.~\textit{fat jet}. Recent
proposals for finding a boosted Higgs boson decaying into a $b \bar b$
pair are based on the investigation of the substructure of each fat
jet~\cite{Butterworth:2008iy,Abdesselam:2010pt}. Within these
approaches the Higgs boson candidate is a multi-jet system which
should contain not only the Higgs boson decay products, but also QCD
radiation associated to them. It is therefore extremely important to
have predictions for the $VH$ process that implement gluon radiation,
both from the initial and final state.

Higher-order corrections to the Higgsstrahlung process, with the Higgs
boson decaying into a $b\bar b$ pair, have been known since a long
time. NLO corrections to Higgs boson production in association with a
vector boson have been computed in
Refs.~\cite{Hamberg:1990np,Baer:1992vx,Ohnemus:1992bd} and implemented
in the program MCFM~\cite{mcfm}. Leading electro-weak corrections are
available as well~\cite{Ciccolini:2003jy} and implemented, together
with QCD corrections, in the program HAWK~\cite{Denner:2011id}. NNLO
results exists for the total cross section~\cite{Brein:2003wg} for
both $WH$ and $ZH$ processes, while a fully differential code is
available for $WH$ production only~\cite{Ferrera:2011bk}.  In all
these production codes the decay of the Higgs into a $b\bar b$ pair is
implemented at LO only. Concerning decay, NLO corrections for massive
bottom quarks have been computed in Ref.~\cite{Drees:1990dq}, while
for massless bottom quarks a fully differential calculation is
available at NNLO~\cite{Anastasiou:2011qx}. NLO corrections have been
interfaced to parton showers in the MC@NLO framework in
Ref.~\cite{LatundeDada:2009rr}. Furthermore, starting from version
2.5, Herwig++~\cite{Gieseke:2011na} implements NLO corrections to
Higgs boson production~\cite{Hamilton:2009za} and Higgs boson decay,
both matched independently to parton shower.

Investigating the effect of NLO QCD corrections to both production and
decay on present and future Higgs searches is the aim of the present
Letter. We do this in the case of Higgs boson production in association
with a $W$ boson. Since to our knowledge no fixed-order program
implements NLO QCD corrections to both Higgs boson production and
decay, we have decided to construct a new code based on the already
available matrix elements. Having our own fully differential code give
us the opportunity to study NLO QCD corrections to production and
decay in a completely separate way. We stick to a fixed-order
calculation since its outcome can be interpreted more easily than
the corresponding one from Monte Carlo event generators. The latter in
fact supplements the hard matrix element with soft and/or collinear
emissions, which can induce modifications on pure NLO
effects. However, although we do not present any prediction obtained
with Monte Carlo event generators, we will discuss how potential
instabilities of fixed-order calculations can be removed either with
all-order QCD resummation or with parton shower Monte Carlo's.

We now describe the details of our calculation.  Denoting by
$d\sigma_{pp\to WH}$ the differential cross section for $WH$
production and by $d\Gamma_{H \to b\bar b}$ the differential decay
rate for a Higgs boson decaying into a $b\bar b$ pair we have the
perturbative expansions
\begin{equation}
  \label{eq:2}
  d\sigma_{pp\to WH} = d\sigma_{pp\to WH}^{(0)}+d\sigma_{pp\to
    WH}^{(1)}\,,\qquad
d\Gamma_{H \to b\bar b} = d\Gamma_{H \to b\bar b}^{(0)} + d\Gamma_{H \to b\bar b}^{(1)}\,,
\end{equation}
where $d\sigma_{pp\to WH}^{(1)}$ is of relative order $\alpha_s$ with
respect to $d\sigma_{pp\to WH}^{(0)}$ (and similarly $ d\Gamma_{H \to
  b\bar b}^{(1)}$ with respect to $ d\Gamma_{H \to b\bar
  b}^{(0)}$). Using the narrow width approximation, which is reasonable
for a light SM Higgs, we can combine NLO corrections to
production and decay as follows
\begin{multline}
  \label{eq:3}
  d\sigma_{pp\to (H \to b\bar b) W} = \left(d\sigma_{pp\to
      WH}^{(0)}\times \frac{d\Gamma_{H \to b\bar b}^{(0)} + d\Gamma_{H \to b\bar b}^{(1)}}{\Gamma_{H \to b\bar b}^{(0)} + \Gamma_{H \to b\bar b}^{(1)}}
+d\sigma_{pp\to
    WH}^{(1)} \times \frac{d\Gamma_{H \to b\bar b}^{(0)}}{\Gamma_{H \to b\bar b}^{(0)}}\right) \times \mathrm{Br}(H\to b\bar b)\,,
\end{multline}
where $\Gamma_{H \to b\bar b}^{(0)}$ is the LO total $H\to b \bar b $
decay rate, $\Gamma_{H \to b\bar b}^{(1)}$ the corresponding NLO
correction and $\mathrm{Br}(H\to b\bar b)$ is the branching ratio for
the decay $H\to b\bar b$. Before describing the phenomenology we
briefly give some details of the calculation. To handle infrared
divergences we use a fully local subtraction method both for
production and decay. We first compute real and virtual matrix
elements in $4-2 \epsilon$ dimensions. We then suitably parametrise
the phase space for the emission of a single real gluon, expand each
denominator occurring in the real matrix element in powers of
$\epsilon$, and cancel all the resulting $1/\epsilon^2$ and
$1/\epsilon$ poles point by point in phase space either against
virtual corrections, or against the collinear counterterm provided by
the $\overline{\rm MS}$ factorisation scheme. For the production
process, we use, depending on the computation order, the LO or NLO
MSTW2008 parton densities~\cite{Martin:2009iq} interfaced through
LHAPDF~\cite{Whalley:2005nh}, the latter corresponding to
$\alpha_s(M_Z)=0.120179$, which is the value we use for NLO
corrections. We consider both $W^+$ and $W^-$ production, keeping the
full spin correlations when letting them decay into a lepton and a
neutrino. Concerning Higgs boson decay, all matrix elements for $H \to
b \bar b$ are computed for massive bottom quarks, using the on-shell
renormalisation scheme with a pole mass $m_b = 4.24\GeV$. We have
checked that our production code agrees with MCFM, and our total decay
rate reproduces the NLO result of
Ref.~\cite{Drees:1990dq}. Furthermore, in the following we will
consider the production of a Standard Model Higgs boson of mass
$m_H=125\GeV$, with $\mathrm{Br}(H\to b\bar b)=0.578$ taken from
Refs.~\cite{Djouadi:1995gt,Djouadi:1997yw,Dittmaier:2011ti}.\footnote{In
  fact, since we will be concerned mainly on $K$-factors and shapes of
  distributions, the actual value of $\mathrm{Br}(H\to b\bar b)$ will
  not be relevant for the main issues discussed in the Letter.}

\section{NLO corrections to Higgs searches with the fat-jet method}
\label{sec:boost}

A natural place to look for a boosted Higgs boson is the LHC with
$\sqrt s = 14\TeV$ at high luminosity, where one has the possibility
to cut on a high-transverse momentum Higgs boson, and still have a
number of events that make it possible to significantly distinguish the
signal from the background. Therefore, we first give theoretical
predictions for observables that are of use at the LHC with $\sqrt s =
14\TeV$ when searching for a boosted Higgs boson associated to a $W$
boson, using the strategy of Ref.~\cite{Butterworth:2008iy}.  In the
following we describe the set of kinematical cuts we employ for our
theoretical analysis. First of all, we put some basic constraints on
the decay products of the $W$ boson, namely that the charged lepton
has a transverse momentum $p_T^l > 30\GeV$ and a pseudorapidity
$|\eta_l| < 2.5$, and that the total missing transverse momentum
fulfils $p_T^{\rm miss}>30\GeV$. We then require that the
reconstructed $W$ boson have large transverse momentum $p_T^W >
210\GeV$. This value is approximately equal to the minimum transverse
momentum that a boosted Higgs boson recoiling against a $W$ boson must
have, at tree level, so that the $b\bar b$ pair resulting from its
decay falls into a cone of radius $R=1.2$. The latter is the value of
jet radius $R$ considered in
Ref.~\cite{Butterworth:2008iy}. Specifically, a Higgs boson decaying
into a $b\bar b$ pair is searched for by clustering each event into
fat jets using the Cambridge/Aachen
algorithm~\cite{Dokshitzer:1997in,Wobisch:1998wt} from the software
package FastJet~\cite{Cacciari:2011ma} with $R=1.2$, and examining the
substructure of each jet to see if it contains the Higgs boson decay
products. Once we have identified a fat jet, in order to establish
whether it can be considered a Higgs candidate, we follow the
procedure proposed in Ref.~\cite{Butterworth:2008iy}, which we briefly
recall:
\begin{enumerate}
\item we undo the last clustering inside the fat jet $j$, thus identifying
  two subjets $j_1$ and $j_2$ ordered according to their invariant mass, $m_{j_1}^2 > m_{j_2}^2$;
\item we require a significant mass drop $m_{j_1}^2 < \mu m_j^2$ and
  impose $\max\{p_{T,1}^2,p_{T,2}^2\} \Delta R^2_{j_1,j_2} > y_{\rm
    cut}\> m_j^2$, in order to suppress asymmetric splittings. 
\end{enumerate}
If both conditions are fulfilled then $j$ is a candidate Higgs jet and
the procedure terminates. Otherwise $j=j_1$ and we go back to step 1.
The fat jet is then kept as a Higgs candidate only if both $j_1$ and
$j_2$ have $b$-tags. Finally, again following
Ref.~\cite{Butterworth:2008iy}, one should apply a filtering
procedure, which consists in reclustering the candidate Higgs jet
with a radius $R_{\rm filt} < R$, and then choosing the candidate
Higgs mass as the invariant mass of the hardest (i.e.\ with the
highest $p_T$) $n_{\rm filt}$ subjets. 
Since our calculation is pure NLO for both production and
decay, a fat jet will contain at most three subjets. As suggested in
Ref.~\cite{Butterworth:2008iy}, we choose $n_{\rm filt}=3$, and therefore we
can skip the filtering step at this stage. However the full procedure
has been programmed in our numerical code, so that it can be used when
NNLO corrections to both production and decay will be implemented.

\begin{figure}[ht]
  \centering
  \includegraphics[width=\textwidth]{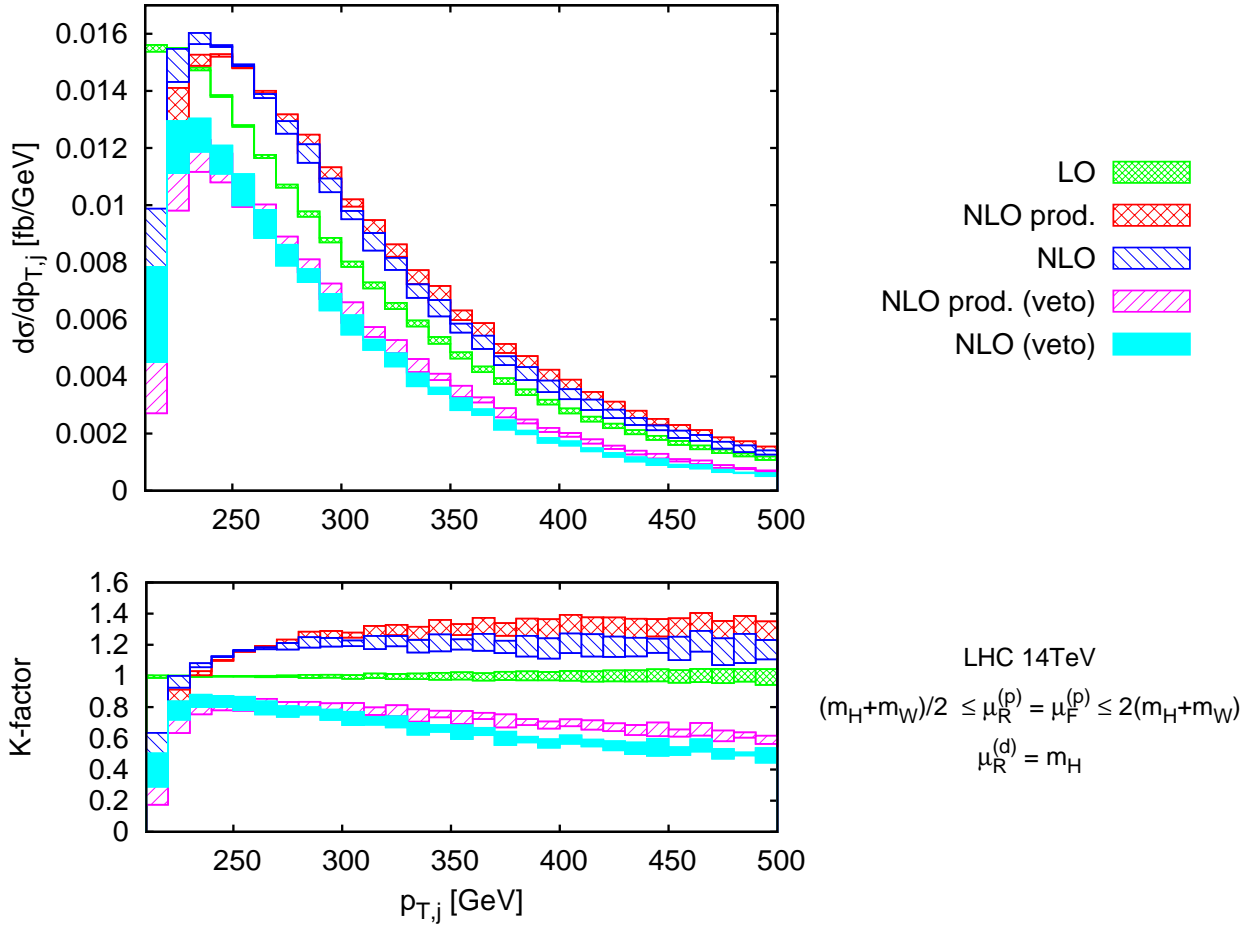}
  \caption{The transverse momentum distribution of the candidate Higgs
    fat jet at the LHC with $\sqrt s = 14\TeV$, corresponding to the
    kinematical cuts described in the text. (For
interpretation of the references to colour, the reader is referred to the web version of this Letter.)}
  \label{fig:ptj}
\end{figure}

The first relevant observable we consider is the transverse momentum
$p_{T,j}$ of the candidate Higgs jet. In particular, we wish to
perform an analogous study as in Ref.~\cite{Ferrera:2011bk}, including
NLO corrections to both Higgs boson production and decay. As in
Ref.~\cite{Ferrera:2011bk}, we require the candidate Higgs jet to be
the one with the highest transverse momentum and to be central,
$|\eta_j|<2.5$. In Fig.~\ref{fig:ptj} we show plots for the fat-jet
$p_T$-spectrum corresponding to two different event selection
procedures. In the first case no constraint is imposed on any extra
jet, whilst in the second case we impose a jet-veto condition,
requiring that there are no further jets with $p_T>p_{T,{\rm
    veto}}=30\GeV$ and $|\eta| < \eta_{\rm veto}=3$ (again according
to what is done in Ref.~\cite{Butterworth:2008iy}). The perturbative
stability of our predictions is investigated by simultaneous variation
of renormalisation and factorisation scales for the production process
by a factor of two around $\mu^{(p)}_{R}=\mu^{(p)}_{F}=m_H+m_W$. Since
we know from the study of Ref.~\cite{Rubin:2010fc} that, for the
fat-jet analysis described above, no infrared problems are expected
from QCD corrections to the decay process, we have decided to fix the
renormalisation scale for the decay at $\mu^{(d)}_R=m_H$.

{
\renewcommand{\arraystretch}{1.3}
\begin{table}[h]
  \centering
  \begin{tabular}{|l|c|c|c|}
    \hline
     &LO& NLO (prod.) & NLO \\  \hline 
    $\sigma_{\rm inc}\>[\fb]$     & $1.53^{+0.02}_{-0.03}$ & $1.87^{+0.05}_{-0.05}$ & $1.80^{+0.06}_{-0.07}$ \\ 
    \hline
    $\sigma_{{\rm 0-jet}}\>[\fb]$ & $1.53^{+0.02}_{-0.03}$ & $1.19^{+0.04}_{-0.06}$ & $1.12^{+0.06}_{-0.08}$ \\   
    \hline
  \end{tabular}
  \caption{Cross section for Higgs boson production in association to a high-$p_T$ $W$ boson selected according to the cuts described in the main text, for $230\GeV<p_{T,j}<500\GeV$, with ($\sigma_{\rm 0-jet}$) and without ($\sigma_{\rm inc}$) a veto on an extra jet. The intervals are obtained by varying the scales for production and decay by a factor two around the central values described in the text.
}
  \label{tab:xsct}
\end{table}
}

At LO the distributions with and without an extra-jet veto obviously
coincide.
On the contrary, as observed already in Ref.~\cite{Ferrera:2011bk},
there are substantial differences for NLO distributions. Without a
veto on an extra jet, if one excludes the lowest $p_T$-bin, NLO
corrections to production are positive, giving roughly a constant
$K$-factor of about 1.3 at large $p_{T,j}$. The addition of NLO
corrections to decay does not alter significantly this $K$-factor,
reducing it to 1.2 (see Fig.~1, red and dark-blue bands). The fact
that the $K$-factor is only slightly decreased when adding NLO
corrections to decay suggests that the observable we consider is
sufficiently inclusive with respect to extra gluon radiation from the
$b\bar b$ system. In other words, in the boosted regime we are
considering, final-state QCD radiation is well contained inside the
fat jet, and therefore we observe no large virtual corrections
unbalanced by real radiation. After imposing the jet veto, NLO
corrections become negative, and increase in size when the jet transverse
momentum increases. This is due to virtual contributions that do not
cancel fully against initial-state real radiation, giving a (negative)
logarithmic left-over as large as $\alpha_s \ln^2(p_{T,j}/p_{T,{\rm
    veto}})$.\footnote{Note that double logarithmic contributions
  $\alpha_s \ln^2(p_{T,j}/p_{T,{\rm veto}})$ occur in this case
  because, as explained in Ref.~\cite{Banfi:2004yd},
  $\ln(p_{T,j}/p_{T,{\rm veto}})$ is smaller than the maximum rapidity
  $\eta_{\rm veto}$ available for the additional jets. These
  logarithms could in principle be resummed at all orders with the
  methods developed in Refs.~\cite{Banfi:2004yd,Banfi:2012yh,Banfi:2012jm,Becher:2012qa}.} We
observe that, also in this case, the addition of NLO corrections to
decay causes only a mild reduction of the $K$-factor, around 10\% and
roughly constant over the whole fat-jet $p_T$-range (see Fig.~1, purple
and light-blue bands).
An important remark is in order concerning the behaviour of the
fat-jet $p_T$-spectrum in the lowest $p_T$-bin. There one notices a
significant decrease in the cross section in going from LO to NLO, as
well as a larger variation when changing renormalisation and
factorisation scales. This bin corresponds in fact to the situation in
which one imposes symmetric $p_T$ cuts on both the Higgs boson and the
$W$ boson. As observed in Refs.~\cite{Klasen:1996yk,Frixione:1997ks}
and explained in Ref.~\cite{Banfi:2003jj}, symmetric cuts can cause
instabilities in the QCD perturbative
series. However~\cite{Banfi:2003jj}, such instabilities could be
removed by performing a resummation of large logarithms appearing in
the distribution in the transverse momentum of the $HW$ system. The
physics underlying such resummation is implemented in all parton
shower Monte Carlo's, which should then be used for Higgs
searches including the symmetric-cut region. Performing such a
resummation is beyond the scope of this work. Therefore, we restrict
our subsequent analyses to the asymmetric-cut region $p_{T,j} >
230\GeV$, where our fixed-order predictions seem to be reliable.
Finally, since our code is not public yet, in Table~\ref{tab:xsct} we
report an example of the cross sections we obtain for the integrated
$p_T$ spectrum for $\mu_R^{(p)}=\mu_F^{(p)}=m_H+m_W$ and of
$\mu_R^{(d)}=m_H$, together with the corresponding renormalisation and
factorisation scale uncertainties.

\begin{figure}[htbp]
  \centering
  \includegraphics[width=0.9\textwidth]{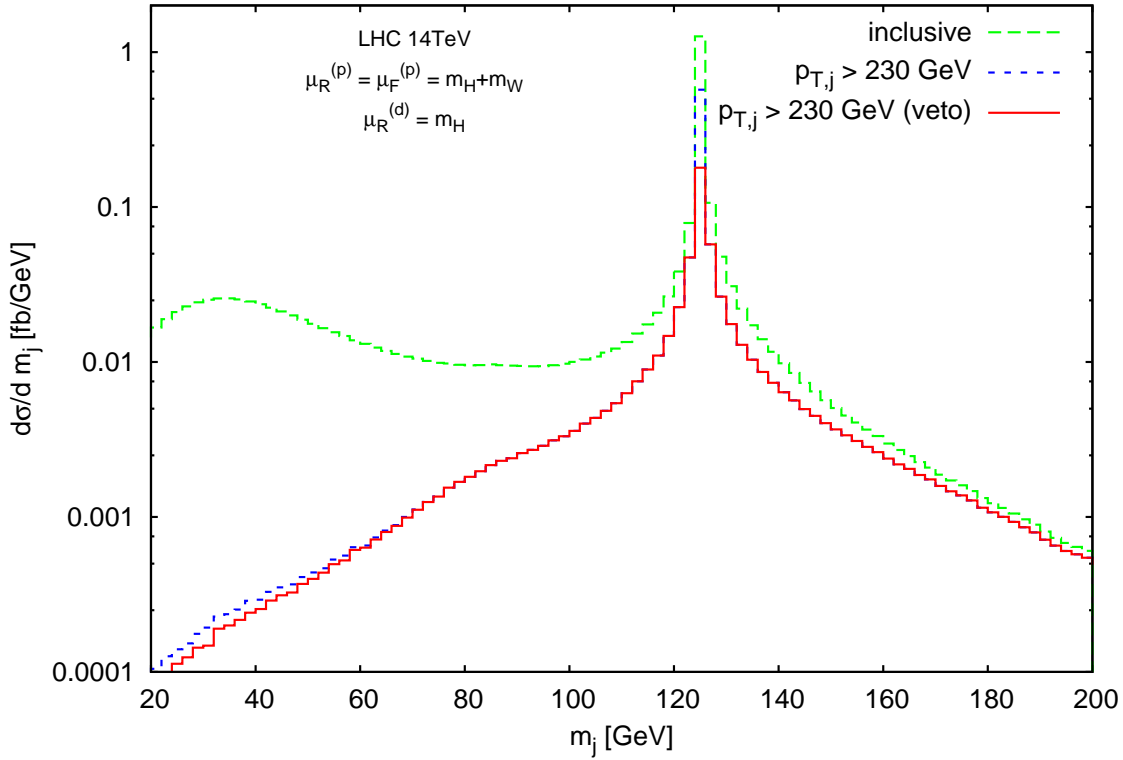}
  \caption{The distribution in the invariant mass of the candidate
    Higgs jet, without any kinematical cuts (green, dashed, labelled
    ``inclusive''), fully inclusive with respect to all other jets
    (blue, dashed, labelled ``$p_{T,j}>230 \GeV$''), and with a
    jet-veto condition (red, solid, labelled ``$p_{T,j}>230 \GeV$ (veto)''). All
    curves correspond to $\mu_R^{(p)}=\mu_F^{(p)}=m_H+m_W$ for Higgs
    boson production and $\mu_R^{(d)}=m_H$ for its decay.}
  \label{fig:mass}
\end{figure}

From the plots in Fig.~\ref{fig:ptj} it seems that the net effect of
QCD corrections to Higgs boson decay is just that of reducing the
production rate of a candidate Higgs fat jet. However, the fat jet
considered there can have an arbitrary invariant mass, whilst in
general one measures the fat-jet invariant mass distribution, given a
set of kinematical cuts, and looks for a mass peak. Therefore, it is
useful to investigate the impact of NLO corrections to Higgs boson
production and decay over the reconstruction of a mass peak based on
the fat-jet analysis described at the beginning of this section. At
NLO there are two effects that can spoil this reconstruction, both
triggered by gluon radiation. The first is the emission of a parton
from the initial state that is subsequently clustered within the fat
jet. The second is the loss of gluon radiation from the $b\bar b$ pair
originating from the decay of the Higgs boson. Both effects are
studied through the plots in Fig.~\ref{fig:mass}, showing the
differential distribution $d\sigma/dm_j$ in the invariant mass of the
fat jet, for $\mu_R^{(p)}=\mu_F^{(p)}=m_H+m_W$ and
$\mu_R^{(d)}=m_H$. 
The first NLO curve in Fig.~\ref{fig:mass} (green, dashed) corresponds
to the fully inclusive situation in which the only selection
requirement is that there is a candidate Higgs jet, with no cut
whatsoever on the jet transverse momentum. This curve is shown to
illustrate how the fat-jet selection technique works in practice.  We
first observe that the fat-jet method is pretty robust under radiative
corrections, in that, even without requiring a high-$p_T$ $W$ boson,
only around 30\% of candidate Higgs events fall outside the mass
window $110\GeV< m_j < 140\GeV$ (a typical bin size for boosted Higgs
searches at the LHC, see~\cite{Chatrchyan:2012ww}). The region to the
right of the peak corresponds to situations in which initial-state
radiation is clustered inside the fat jet, thus artificially
increasing the invariant mass of the latter. To the left of the peak
we see a long tail corresponding to events in which a gluon emitted
from the $b\bar b$ system originating from Higgs boson decay escapes
the fat jet. This effect is entirely due to NLO corrections to Higgs
boson decay, and its contribution to degrading the resolution of the
mass peak is comparable to that coming from NLO corrections to Higgs
boson production. In fact, most events outside the mass window
$110\GeV< m_j < 140\GeV$ have $m_j<110\GeV$. These events extend down
to $m_j = 2m_b$, corresponding to the situation in which the $b\bar b$
pair recoils against a hard gluon. The other two curves correspond to
events passing the same kinematical cuts as in Fig.~\ref{fig:ptj}, and
with an additional cut on the fat-jet transverse momentum
$p_{T,j}>230\GeV$. We first observe that this cut reduces considerably
the fraction of events with $m_j<110\GeV$ which is around 10\% both
with and without the jet veto. A further remark is in order concerning
the curve (solid, red) obtained by imposing the additional constraint
of vetoing all extra jets with $p_T> 30\GeV$ and rapidity
$|\eta|<3$. In this case, as expected from the study of
Ref.~\cite{Ferrera:2011bk} and Fig.~\ref{fig:ptj}, the peak height is
reduced due the suppression of real emission and the dominance of
negative uncancelled virtual corrections.  Among the effects of the
jet veto there is also that of eliminating events in which a gluon
emitted by the $b\bar b$ system escapes the fat jet. For the red solid
curve this is visible in the figure for $m_j < 60\GeV$ and has a
negligible impact on the resolution of the mass peak. We finally
remark that the fact that the mass peak survives depends crucially on
both the jet-veto condition and the procedure used to identify a
candidate Higgs jet. We have observed that decreasing $p_{T,{\rm
    veto}}$ can lower the number of selected events in such a way that
a peak is not visible any more. The same remark holds for alternative
procedures for defining a candidate Higgs jet, which need to be tested
against final-state QCD radiation as well. An example of how the
latter can affect significantly the outcome of a Higgs search analysis
is discussed in the next section.

\section{Higgs searches at the LHC with $\mathbf{\sqrt s =
    8}$~TeV}
\label{sec:lhc}

With the current LHC energy, CMS is searching for a high-$p_T$ Higgs
radiated off a vector boson. They do not perform a full fat-jet
analysis, but instead use a simplified procedure aimed at identifying
a boosted hadronic system that could be considered as a Higgs
candidate~\cite{Chatrchyan:2012ww}. Here is a summary of the CMS
analysis. First, they impose cuts on the decay products of the $W$
boson. For a $W$ boson decaying into a muon and its associated
neutrino (the case we consider in the following), they require $p_T^l
>20\GeV$ and $|\eta_l|<2.4$, together with a constraint on the missing
transverse energy $p_T^{\rm miss} >35\GeV$. The Higgs candidate is a
dijet system consisting of two central ($|\eta|<2.5$) $b$-tagged jets
with $p_T>30\GeV$, reconstructed with the anti-$k_t$
algorithm~\cite{Cacciari:2008gp} with $R=0.5$. Then, high-$p_T$ events
are selected by imposing a cut both on the transverse momentum of the
reconstructed $W$ boson $p_T^W>160\GeV$ and on that of the dijet
system $p_{T,j}>165\GeV$, and requiring the latter to be central
($|\eta_{j}|<2.5$). Finally, the $W$ boson and the Higgs candidate are
required to be almost back-to-back in the transverse plane, by
imposing $\Delta \phi_{W,j}\equiv|\phi_W-\phi_{j}|>3$, and no extra
jets are allowed with $p_T>20\GeV$ and $|\eta|<2.4$.

\begin{figure}[ht] 
  \centering
  \includegraphics[width=0.7\textwidth]{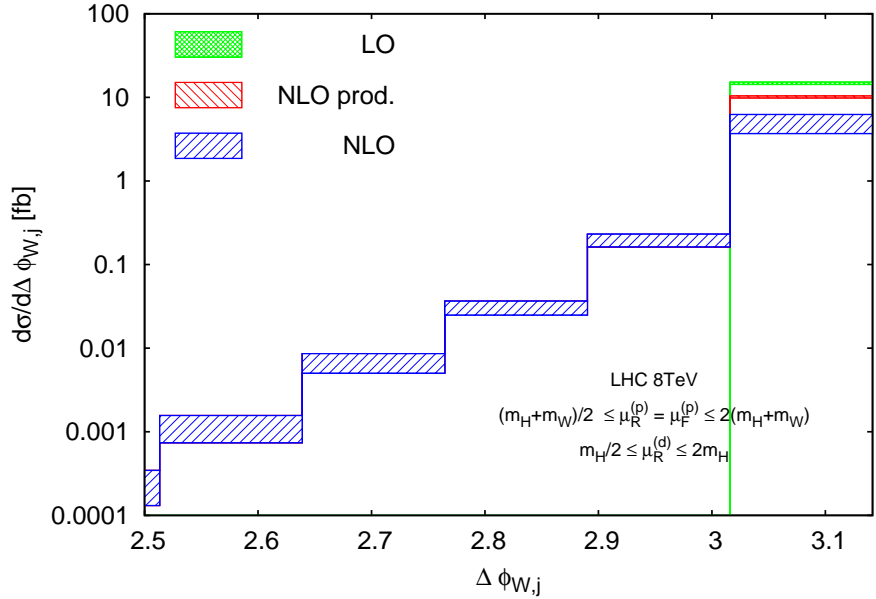}
  \caption{The differential distribution in the azimuthal angle
    $\Delta \phi_{W,j}$ between the reconstructed $W$ boson and the candidate Higgs jet,
    corresponding to the selection cuts described in the text.
}
  \label{fig:deltaphi}
\end{figure}
Among these conditions, the requirement on $\Delta \phi_{W,j}$, is
particularly sensitive to initial-state radiation, in particular soft
and collinear gluon emissions along the beam. We wish therefore to
investigate if our predictions for the $\Delta \phi_{W,j}$
distribution are stable against higher-order corrections. We do this
via simultaneous variations of renormalisation and factorisation
scales for the production process around
$\mu_R^{(p)}=\mu_F^{(p)}=m_H+m_W$, and independent variation of the
renormalisation scale for the decay around $\mu_R^{(d)} =
m_H$. Fig.~\ref{fig:deltaphi} shows the $\Delta \phi_{W,j}$
distribution, obtained after imposing the cuts described above. First
of all, one notices that most events are concentrated in the bin
$\Delta \phi_{W,j} > 3$, so that we expect that bin to be wide enough
to ensure a sufficiently inclusive cancellation of large real and
virtual corrections arising from the region close to $\Delta
\phi_{W,j} = \pi$. This is confirmed by the fact that, if one
considers production only (the histogram labelled ``NLO prod.''), the
$K$-factor we observe in that bin is moderate (around 0.8). Indeed, the
cut on $\Delta \phi_{W,j}$ is only one of the effects that are
responsible for such $K$-factor, the others being the cut on the jet
$p_T$ and the jet-veto condition, which can also put constraints on
initial-state radiation. We remark that, if the constraint on $\Delta
\phi_{W,j}$ were moved closer to $\pi$ while keeping all other cuts
fixed, one would expect large logarithmic contributions $\alpha_s^n
\ln^m(\pi-\Delta \phi_{W,j})$ arising from multiple initial-state
soft-collinear emissions. These could be resummed, either analytically,
or using Monte Carlo event generators. When adding final-state
radiation, one observes that the height of the distribution in the
rightmost bin is further depleted. However, due to the fact that, for
$\Delta \phi_{W,j} < 3$ all NLO distributions basically coincide, this
depletion cannot be ascribed to a restriction on final-state radiation
imposed indirectly through the cut on $\Delta \phi_{W,j}$. The
reduction of the cross section is mainly due to the loss of QCD
radiation from the $b \bar b$ system. Due to the jet-veto condition,
any gluon that is not clustered inside the two $b$-jets that
constitute the Higgs candidate is likely to be soft, and therefore
$\Delta \phi_{W,j}$ will be close to $\pi$. It is this restriction on
final-state radiation that causes an imbalance between real and
virtual corrections, giving a large negative contribution. A further
source of large virtual corrections, which contribute significantly to
the size of the observed $K$-factor, is the presence of a term $\alpha_s
\ln(m_b/m_H)$ in the virtual corrections, coming from the on-shell
renormalisation of the coupling of the $b$ quark to the Higgs. To
investigate the impact of this term, we use a different prescription
to combine NLO corrections to production and decay, by strictly
expanding Eq.~(\ref{eq:3}) at order $\alpha_s$:
\begin{multline}
  \label{eq:4}
  d\sigma^{\rm exp}_{pp\to (H \to b\bar b) W} = \left\{d\sigma_{pp\to
      WH}^{(0)}\times \left[\frac{d\Gamma_{H \to b\bar b}^{(0)}}{\Gamma_{H \to b\bar b}^{(0)}}\left(1-\frac{\Gamma_{H \to b\bar b}^{(1)}}{\Gamma_{H \to b\bar b}^{(0)}}\right) + \frac{d\Gamma_{H \to b\bar b}^{(1)}}{\Gamma_{H \to b\bar b}^{(0)}}\right]
\right. \\ \left. + d\sigma_{pp\to
    WH}^{(1)} \times \frac{d\Gamma_{H \to b\bar b}^{(0)}}{\Gamma_{H
      \to b\bar b}^{(0)}}\right\} \times \mathrm{Br}(H\to b\bar b)\,.
\end{multline}
We see that the curve corresponding to this last prescription (purple,
labelled ``NLO exp.'') is significantly higher than the curve
corresponding to Eq.~(\ref{eq:3}) (blue, labelled ``NLO''). The
difference between the two prescriptions can be understood as an
indication on the convergence of perturbation theory. In this respect,
we have checked that all the results we have obtained in the previous
section using the fat-jet procedure are, within scale uncertainties,
insensitive to a change of prescription, thus indicating that this
procedure is inclusive enough with respect to final-state radiation to
ensure good convergence of the perturbative expansion. In the
following, whenever relevant, we will use both prescriptions.

We now proceed by presenting the distributions studied in the previous
section, this time relative to the
candidate Higgs selected according to the CMS procedure, and for
LHC at $\sqrt s = 8 \TeV$.\footnote{We have checked that our
  considerations do not change in the case of LHC at $\sqrt s =
  7\TeV$.} For the mass distribution, we will also compare the mass
spectrum corresponding to the CMS analysis with that obtained with
the fat-jet analysis described at the beginning of
section~\ref{sec:boost}. In this case $m_{j}$ will label the
invariant mass of the fat jet.
\begin{figure}[ht]
  \centering
  \includegraphics[width=1.0\textwidth]{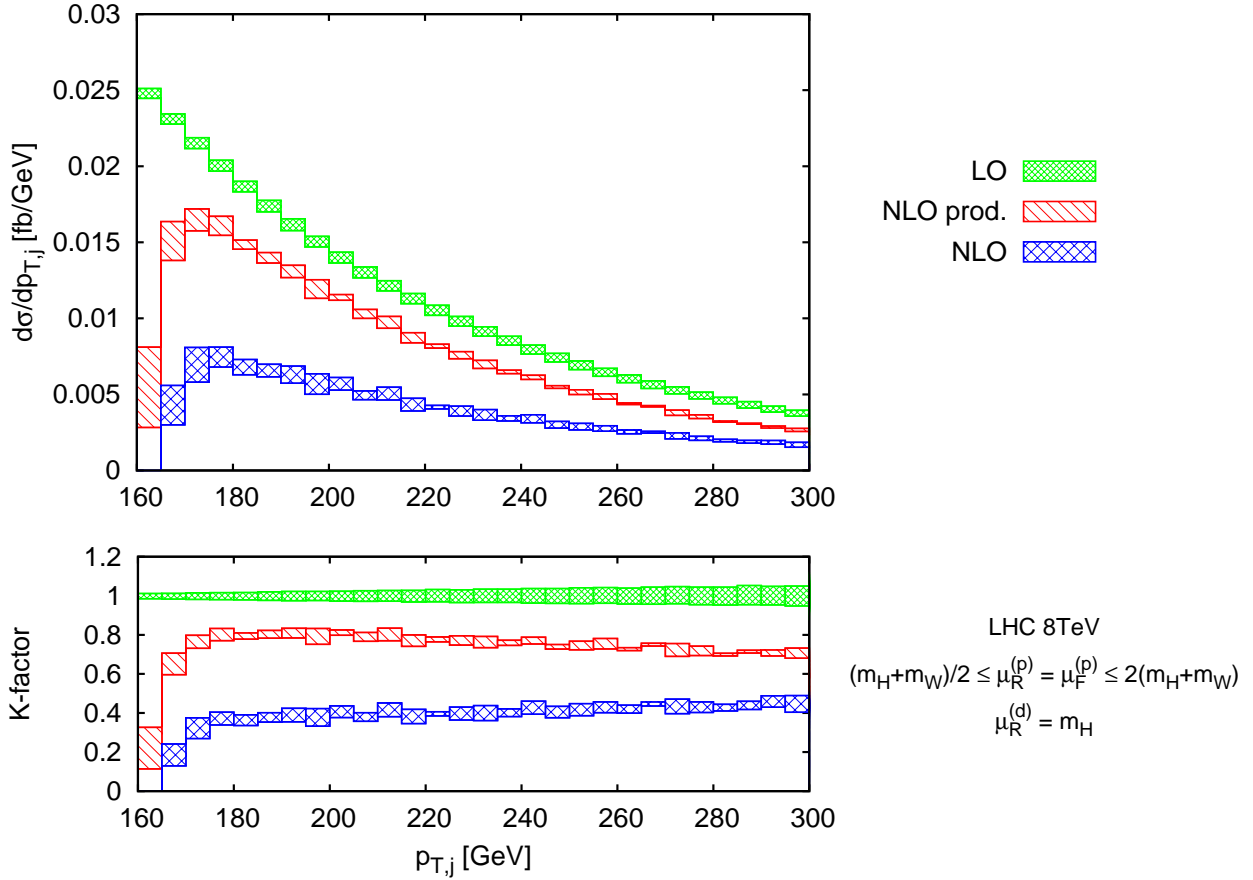}
  \caption{The distribution in the transverse momentum $p_{T,j}$ of
    the candidate Higgs dijet system, corresponding to the selection
    cuts described in the text.}
  \label{fig:ptjet_cms}
\end{figure}

Fig.~\ref{fig:ptjet_cms} contains distributions in the transverse
momentum $p_{T,j}$ of the candidate Higgs dijet system. Each band
corresponds to a simultaneous variation of renormalisation and
factorisation scale by a factor of two around
$\mu_R^{(p)}=\mu_F^{(p)}=m_H+m_W$ in the production process, while
renormalisation scale for the decay is kept fixed at
$\mu_R^{(d)}=m_H$. If one does not include NLO corrections to decay,
one observes a 20\% reduction in the cross section with respect to
LO. This reduction is driven mainly by the jet-veto condition, and as
expected gets more important as the dijet transverse momentum
increases. The inclusion of NLO corrections to Higgs boson decay causes
a further decrease of the cross section.  We interpret this result as
a sizable loss of QCD final-state radiation by the two jets that
constitute the Higgs candidate system, likely due to the fact that the
$b$ jets have a small radius and the typical perturbative jet
$p_T$-loss increases with decreasing radius, as explained in
Ref.~\cite{Dasgupta:2007wa}, and the lost radiation undergoes a
jet-veto constraint. This in turns causes a poor convergence of the
perturbative series, as one observes by comparing the ``NLO'' and
``NLO exp.'' curves. We have checked that this does not happen if one
performs a fat-jet analysis with the same parameters as in
Sec.~\ref{sec:boost}, where one observes instead a $K$-factor of
around 0.6 for $p_{T,j} > 250\GeV$, although in this case the
$p_{T,j}$ distribution drops abruptly, as expected, for $p_{T,j} <
200\GeV$. We have also checked that for larger $p_T$ values
($p_{T,j}>350\GeV$) the CMS and fat-jet procedure give comparable
$p_{T,j}$ spectra. Another remark concerns the first bin ($160 \GeV <
p_{T,j} < 165\GeV$), where the distribution is unstable against scale
variations (it becomes even negative if one includes corrections to
Higgs boson decay), again due to the fact that this bin corresponds to
symmetric transverse momentum cuts, in which the $W$ boson and the
$b$-dijet system recoil against a soft-collinear gluon. This bin is
not included in the CMS analysis, and will not be considered in all
our subsequent studies.

\begin{figure}[ht]
  \centering
  \includegraphics[width=0.8\textwidth]{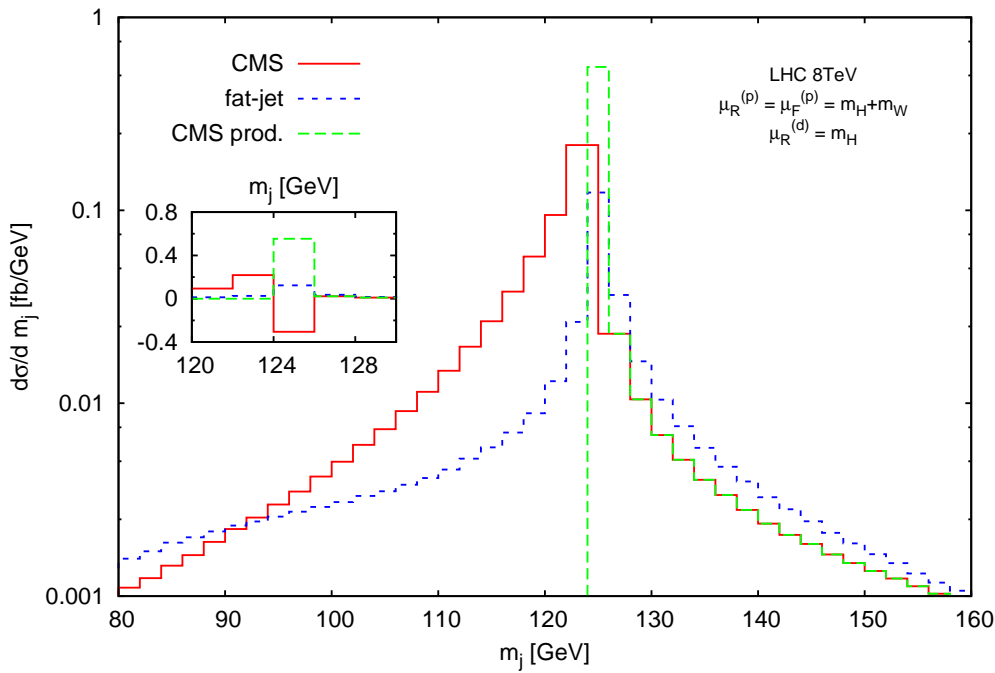}
  \caption{The distribution in the invariant mass $m_j$ of the
    candidate Higgs jet obtained with the procedure adopted by CMS,
    with and without NLO corrections to Higgs decay, and with the
    fat-jet procedure described in Sec.~\ref{sec:boost}.}
  \label{fig:mass_cms}
\end{figure}

We then present in Fig.~\ref{fig:mass_cms} the distribution in the
invariant mass of the candidate Higgs system $d\sigma/dm_{j}$. We
consider here four distributions, the first (solid, red) obtained with
the CMS selection procedure, the second (dashed, blue) corresponding
to the fat-jet selection procedure explained at the beginning of
Sec.~\ref{sec:boost}, with the same selection cuts as CMS for the
leptons, the $W$ boson and the candidate Higgs system, the third
(dashed, green) corresponding to the CMS procedure but where only NLO
corrections to the production process are considered, and the fourth
(dotted, purple) again corresponding the CMS procedure and obtained
using Eq.~(\ref{eq:4}).
From the plots we see that the mass distribution resulting from the
CMS procedure (red, solid), catches more candidate Higgs events than
the fat-jet one, as expected due to the lower $p_T$-cut on the
selected $b$-jets. However, the mass distribution does not display a
mass peak in the expected position. In fact the value of the
distribution at $m_{j} = 125\GeV$ is negative (see inset plot of
Fig.~\ref{fig:mass_cms}) if one uses Eq.~(\ref{eq:3}), and only
slightly positive if one uses instead Eq.~(\ref{eq:4}) (the purple
dotted curve labelled ``CMS exp.''). Regardless of the actual value of
the distribution at $m_{j} = 125\GeV$, this clearly indicates that
radiation from the $b\bar b$ pair originating from Higgs boson decay is
not naturally included in the candidate Higgs system. On the contrary,
the fat-jet procedure (blue, dashed) gives correctly a peak at $m_{j}
= 125\GeV$, although with a reduced height with respect to that of the
shifted peak resulting from the CMS procedure. This result does not
change if one uses the alternative prescription of
Eq.~(\ref{eq:4}). The third curve (green, dashed) shows the mass
distribution obtained by considering NLO corrections to production
only. We notice that the peak is in the expected position, with a
height that is roughly five times larger than that of the peak
corresponding to the fat-jet procedure. In this respect, we remark
that the parameters we have chosen for the fat-jet analysis are
identical to those of Sec.~\ref{sec:boost}. In principle one should
redetermine them after a full simulation of signal and background,
including parton shower effects (for instance using the recent
developments of Ref.~\cite{Hamilton:2009za}). This goes beyond the
scope of this work.

\section{Conclusions}
\label{sec:end}

We have implemented NLO corrections to Higgs boson production in
association with leptonically decaying $W$ boson, and to its
subsequent decay into a $b\bar b$ pair, in a numerical code that
returns weighted events, fully differential in the decay products of
the Higgs boson and of the $W$ boson. We have then looked at how NLO
QCD corrections to Higgs boson decay affect various observables that
are relevant for Higgs searches at the LHC. In particular we have
analysed two different experimental setups, one with $\sqrt s=14\TeV$
and the other with $\sqrt s=8\TeV$. In the first case the Higgs boson
is produced in a boosted regime, with its transverse momentum
larger than its mass, and detected using the fat-jet technique
proposed in Ref.~\cite{Butterworth:2008iy}. With our study in
Section~\ref{sec:boost} we assess that the Higgs candidate obtained
with the fat-jet procedure is stable against radiative
corrections. Its mass distribution is peaked at the expected value of
the Higgs boson mass, and the resolution of the peak is reasonably
good (see Fig.~\ref{fig:mass}). We remark that the height of the peak
is sensitive to the jet-veto condition that one imposes on any jet
besides the candidate Higgs fat jet, the stronger the veto the
stronger the suppression of the peak. In our case the height of the
peak obtained after imposing a jet-veto condition is roughly a third
of that we get if we are fully inclusive with respect to all jets. The
second experimental setup we have considered corresponds to what is
done at the moment by the CMS experiment, but for the LHC current
energy $\sqrt s = 8\TeV$. CMS chooses configurations in which both the
Higgs boson and the $W$ boson have high transverse momentum. Then they
do not perform a full fat-jet analysis, but rather consider as a Higgs
candidate a system of two $b$-jets satisfying a set of transverse
momentum and rapidity cuts. Again we have checked how relevant
distributions are influenced by QCD corrections to Higgs boson
decay. We have found that such corrections have a huge impact both on
the candidate Higgs transverse momentum spectrum and on its invariant
mass distribution. In particular, for the latter it turns out that
the effect of an extra jet-veto and the loss of QCD radiation from the
$b\bar b$ system give a displacement of the mass peak from its
expected position, with a poor peak resolution. This suggests an
instability of the CMS procedure against radiative corrections, and
reveals how important it is to have NLO information on the Higgs boson
decay as well. We remark that the use of a parton shower event
generator will give a smoother mass distribution, but will not
significantly improve the mass resolution of the peak, which is mainly
affected by the large $p_T$ loss of the selected $b$-jets. For
comparison we have also studied the jet-mass distribution for a
candidate Higgs jet obtained with the same fat-jet procedure
considered for the LHC at $\sqrt s=14\TeV$. Also in this case the
procedure seems stable under radiative corrections, and we are able to
reconstruct a mass peak in the expected position.

The study we have performed gives some new information on the impact
of higher-order corrections on the Higgsstrahlung process, but is far
from conclusive. More studies are needed both to improve the accuracy
of the calculation, including for instance NNLO corrections to both
production and decay, and to devise procedures to cure the
instabilities we have found in our analysis. We aim to address both
issues in the near future.

\paragraph{Acknowledgements}
We are grateful to Babis Anastasiou for suggesting us to look into
this problem, and for support and illuminating discussions while this
work was being performed. We also thank Gavin Salam for a careful
reading of the manuscript and insightful comments and suggestions. AB
would like also to thank Stefan Gieseke and Keith Hamilton for
discussions on Herwig++, and Stefan Dittmaier for pointing out
relevant references. The work of J.C. is supported by the ERC Starting
Grant for the project ``IterQCD'' and the Swiss National Foundation
under the contract SNF 200020-126632. Part of this work was performed
while A.B. was at ETH Zurich.

\bibliographystyle{jc}
\bibliography{hwdiff}

\end{document}